\documentclass{sig-alternate}

\usepackage[normalem]{ulem}
\usepackage{graphicx}
\usepackage{xspace}
\usepackage{multirow}
\usepackage{array}
\usepackage[table,xcdraw]{xcolor}
\usepackage{amssymb}
\usepackage{tikz}
\usetikzlibrary{patterns}
\usetikzlibrary{positioning, intersections, arrows, shapes.geometric, calc, shapes.callouts, decorations.pathreplacing, decorations.markings, shapes}
\usepackage{pifont}
\usepackage{pgfplots}
\usepgfplotslibrary{statistics}
\usepackage{pgfplotstable}
\pgfplotsset{compat=newest}
\usepackage{filecontents}
\usepackage{authblk}
\usepackage{mathptmx} % This is Times font
\usepackage{fancyhdr}
\usepackage[normalem]{ulem}
\usepackage[hyphens]{url}
\usepackage[sort,nocompress]{cite}
\usepackage[final]{microtype}
\usepackage[keeplastbox]{flushend}
\usepackage[bookmarks=true,breaklinks=true,letterpaper=true,colorlinks,linkcolor=black,citecolor=blue,urlcolor=black]{hyperref}

\newcommand{\CGraph}{Pictor\xspace}
\newcommand{\FIG}{Figure\xspace}
\newcommand{\TAB}{Table\xspace}

\fancypagestyle{firstpage}{
	\fancyhf{}
	
	\fancyhead[C]{\vspace{15pt}\normalsize{ To Appear in IEEE/ACM International Symposium on Microarchitecture, MICRO 2020 }} 
	\fancyfoot[C]{\thepage}
}
\begin{document}

\title{A Benchmarking Framework for Interactive 3D Applications in the Cloud\vspace{-1cm}}
\date{}

\author[1]{Tianyi Liu}
\author[1]{Sen He}
\author[1]{Sunzhou Huang}
\author[1]{Danny Tsang}
\author[2]{Lingjia Tang}
\author[2]{Jason Mars}
\author[1]{Wei Wang}
\affil[1]{The University of Texas at San Antonio}
\affil[1]{\{tianyi.liu, sen.he, sunzhou.huang, danny.tsang, wei.wang\}@utsa.edu}
\affil[2]{University of Michigan}
\affil[2]{\{lingjia, profmars\}@umich.edu}

\maketitle
\thispagestyle{firstpage}
\pagestyle{plain}

\begin{abstract}
With the growing popularity of cloud gaming and cloud virtual reality (VR),
interactive 3D applications have become a major type of workloads for the
cloud. However, despite their growing importance, there is limited public
research on how to design cloud systems to efficiently support these
applications, due to the lack of an open and reliable research infrastructure,
including benchmarks and performance analysis tools. The challenges of
generating human-like inputs under various system/application randomness and
dissecting the performance of complex graphics systems make it very difficult to
design such an infrastructure. In this paper, we present the design of a novel
research infrastructure, \textit{\CGraph}, for cloud 3D applications and
systems. \CGraph employs AI to mimic human interactions with complex 3D
applications. It can also track the processing of user inputs to provide
in-depth performance measurements for the complex software and hardware stack
used for cloud 3D-graphics rendering. With \CGraph, we designed a benchmark
suite with six interactive 3D applications. Performance analyses were conducted
with these benchmarks to characterize 3D applications in the cloud and reveal
new performance bottlenecks. To demonstrate the effectiveness of \CGraph, we
also implemented two optimizations to address two performance bottlenecks
discovered in a state-of-the-art cloud 3D-graphics rendering system, which
improved the frame rate by 57.7\% on average.
\end{abstract}

\section{Introduction}\label{sec:intro}
The rise of cloud gaming and cloud virtual-reality (VR) has made interactive 3D
applications a major type of workloads for cloud computing and data centers~\cite{GeforceNow,Citrix-VDI,Sumerian,Stadia,XCloud}. A main benefit
of rendering interactive 3D applications in the cloud is that it may reduce the
installation and operational costs for the large-scale deployments of these 3D
applications. Running these 3D applications in the cloud also allows mobile
clients with less powerful GPUs to enjoy better visual effects. Moreover, cloud
3D-graphics rendering may also simplify the development and delivery of these 3D
applications. 
For the rest of this paper, we refer to these 3D interactive applications simply
as \textit{3D applications}.

Most prior research on virtual desktop infrastructure (VDI) or cloud gaming
focused on network
latency~\cite{2006-Lai-TOCS,2012-Dusi-COM,2017-Abari-NSDI}. However, the network
latency is considerably reduced today and becomes viable for cloud 3D
applications~\cite{2018-Tan-SIGMETRICS}. This improved network, in turn, makes
the design of \textit{cloud 3D-graphics rendering systems} crucial to the
efficiency and performance of cloud 3D applications. However, there is limited
public research on this system design, largely due to the lack of an open and
reliable research infrastructure, including benchmarks and performance analysis
tools.  Prior attempts to provide such research
infrastructures~\cite{2005-Zeldovich-ATC,2003-Nieh-TOCS,2010-Berryman-CloudCom}
have limited success due to the following challenges.% to achieve high

First, for reliable evaluation, the research infrastructure must be able to
mimic human interactions with 3D applications under randomly generated/placed
objects and varying network latency.
% As the behaviors of interactive 3D applications are heavily affected by human
% interactions, reliably benchmarking these applications requires generating
% human-like inputs. accurately benchmarking their behaviors requires generating
% human-like inputs.  That is, for reliable performance evaluation,
That is, the inputs used for the benchmarks
should closely resemble real human inputs, so that the performance results
obtained with these human-like inputs are similar to those obtained with real
human inputs.
% However, for large-scale evaluation, it is too costly to have many human users
% to interact with the cloud 3D applications. Therefore, benchmarking 3D
% applications requires automatically generating human-like inputs.
Prior research generated human-like inputs from recorded human
actions~\cite{2005-Zeldovich-ATC,2010-Spruijt-LoginVSI}.  However, this
recording does not work for 3D VR applications and games, which have irregular
and randomly placed/generated objects in their frames. Additionally, variations
in network latency may affect when a particular object will be shown on the
screen, further limiting the usefulness of the recorded
actions~\cite{2005-Zeldovich-ATC}.

% during benchmark runs, the frames may arrive at different times than those
% observed at the recording due to varying network and system latency, further
% limiting the usefulness of the recorded
% actions.  %Consequently, accurately simulating human actions requires the benchmarks to recognize randomized GUI objects and issue user inputs properly under any network/system latency.

%\item
Second, to reliably measure performance, the research infrastructure must be
able to accurately measure the round trip time/latency to respond to a user
input~\cite{2003-Nieh-TOCS}, which in turn, relies on the accurate association
of a user input and its response frame. However, this association is very
difficult due to the need to track the handling of a user input and the rending
of its response frame across the network, across the CPU and GPU, and across
multiple software processes (i.e., tracking from the client to the server, and
back to the client).

Third, to effectively identify performance bottlenecks, the research framework
must be able to measure the performance of every stage involved in the handling
of a user input and the rendering of its response frame.
% However, it is very challenging to track the processing of an input (from the
% client to the server, and back to the client), due to the complex graphics
% stack and the heterogeneous
% hardware %and the parallel graphics rendering pipelineand the fact that rendering happens without user input (application automatically update)
% ~\cite{2003-Nieh-TOCS}. Moreover, measuring the performance of every
% processing step also requires inspecting software/hardware components with
% quite different behaviors.
The framework should be able to properly measure the performance of all the
components involved, including those from the complex graphics software stack
and heterogeneous hardware devices.  Additionally, the research framework must
have low overhead to ensure these measurements are reliable.

% \item
Fourth, the research infrastructure should be extensible to easily include new
3D applications. % without changing their source code.
3D applications are typically refreshed every one or two
years %Cloud rendering systems are also likely to be significantly changed due to future research.
and most of them are proprietary. Therefore, the research infrastructure should
be constantly refreshed with new 3D benchmarks without requiring to modify their
source code.
% and cloud systems.
  %To simplify this process, the research infrastructure should be easily extended to new 3D applications 
%\end{enumerate}

In this paper, we present a novel benchmarking framework, called
\textit{\CGraph}, which overcomes the above challenges to allow reliable and
effective performance evaluation of 3D applications and cloud graphics rendering
systems.
% \CGraph is to allow reliable performance
%evaluation for 3D applications and their rendering/managing systems in the
% cloud. % Therefore, \CGraph is designed to generate human-like inputs/actions, so
% that the performance measurements obtained with these human-like actions are the
% same to those obtained with actual human actions. \CGraph is also designed to be
% able to inspect the performance and resource usages of various software/hardware
% components. 
\CGraph has two components: 1) an intelligent client framework that can generate
human-like inputs to interact with 3D applications; and 2) a performance
analysis framework that provides reliable input tracking and
%application-level, system-level and architecture-level
performance measurements. Inspired by autonomous driving, the intelligent client
framework employs computer vision (CV) and recurrent neural network (RNN) to
simulate human actions~\cite{2012-Alex-NIPS,2011-Socher-ICML,1997-Sperduti-NN}.
% The intelligent client framework is designed to ensure that these simulated actions
% are close enough to real human actions, so that performance measurements
% obtained with these simulated actions are the same as those obtained with real
% human actions. %, allowing reliable performance evaluation.
The performance analysis framework tracks inputs with tags and combines various
performance monitoring techniques to measure the processing latency and resource
usage of each hardware and software
component.  %allowing it to be easily applied to new GUI applications.
Additionally, \CGraph is carefully designed to have low overhead and require no
modification to 3D applications.

With \CGraph, we designed a benchmark suite with four computer games and two VR
applications. Through experimental evaluation with these benchmarks, we show
that \CGraph can indeed accurately mimic human action with an average error of
1.6\%. We also conducted an extensive performance analysis on a
state-of-the-art cloud 3D-graphics rendering
system~\cite{1998-Richarson-IC,TurboVNC} to characterize the 3D benchmarks and
the rendering system, analyze the impact of co-locating multiple 3D applications,
and study the overhead of rendering 3D applications in containers.  This
performance analysis demonstrated the benefits of cloud graphics rendering and
revealed new performance bottlenecks. At last, to demonstrate that the in-depth
performance analysis allowed by \CGraph can indeed lead to performance
improvements, we implemented two optimizations which improved average frame-rate
by 57.7\%.

The contributions of this paper include:

1. A novel intelligent client framework that can faithfully mimic human
interactions with complex 3D applications with randomly generated/placed objects
and under varying network latency. 

2. A novel performance analysis framework that can accurately track the
processing of user inputs, and measure the performance of each step involved in
the processing of user inputs in various software and (heterogeneous) hardware
components. This framework is also carefully designed to have low overhead and
require no application source code.

3. A comprehensive performance analysis of a state-of-the-art cloud graphics
rendering system, the 3D benchmarks and containerization. This analysis also
shows the benefit of cloud 3D applications and reveal new optimization
opportunities.

4. Two new optimizations for current cloud graphics rendering system with
significant performance improvements, which also demonstrate the effectiveness
of \CGraph.

The rest of this paper is organized as follows:
Section~\ref{sec:remote_rendering} discusses a typical cloud graphics rendering system;
Section~\ref{sec:benchmark_framework} presents the design of \CGraph;
Section~\ref{sec:evaluation} evaluates the accuracy and overhead of \CGraph;
Section~\ref{sec:perf_analysis} provides the performance analysis on a current cloud graphics rendering system;
Section~\ref{sec:optimization} presents two new optimizations;
Section~\ref{sec:related} discusses related work and Section~\ref{sec:conclusion} concludes the paper.

%%% Local Variables:
%%% mode: latex
%%% TeX-master: "paper"
%%% End:

\section{Cloud 3D Rendering System}\label{sec:remote_rendering}
\begin{figure*}
  \centering
  \includegraphics[width=15cm]{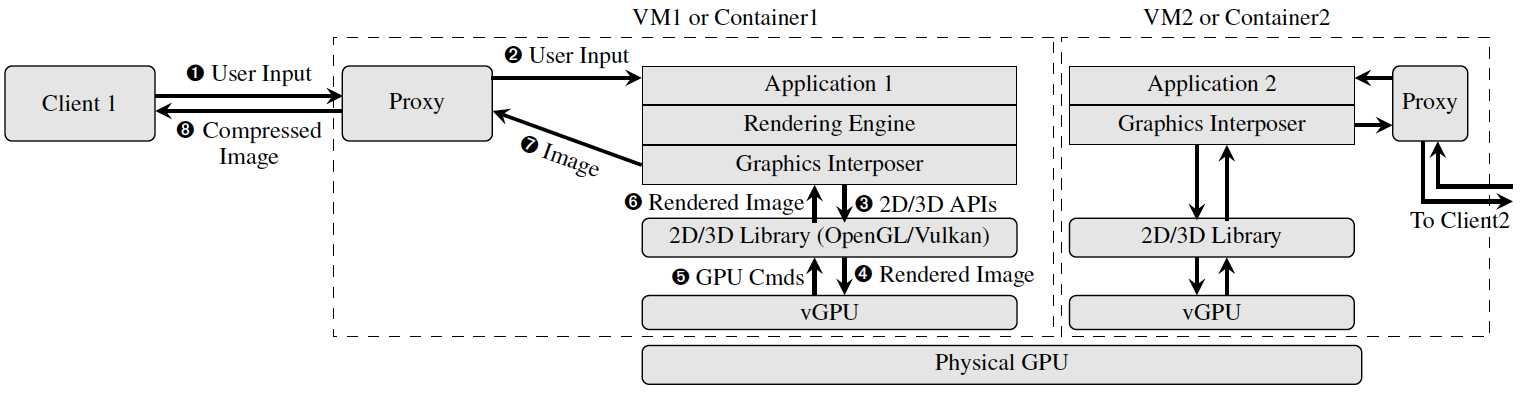}
  \vspace{-2mm}
  \caption{System architecture for cloud graphics rendering.}
  \label{fig:virt_gpu_arch_proxy}
\end{figure*}

Figure~\ref{fig:virt_gpu_arch_proxy} illustrates the typical system architecture
for cloud graphics rendering. This architecture employs a server-client model
where the servers on the cloud execute 3D applications and serve most of their
rendering requests. The client is mainly responsible for displaying UI frames
and capturing user inputs. The client may also perform less-intensive graphics
rendering, depending on the system design.  In this work, we focus on
Linux-based systems and open-source software which are easy to modify and free
to distribute.
% It is possible to design a similar benchmarking framework for Windows-based
% systems, as the system architecture is similar to the one on Linux. However,
% the distribution of the benchmarking framework may be limited by the
% proprietary software. We will consider Windows-based systems in the future.

% Figure~\ref{fig:virt_gpu_arch_proxy} shows the system architecture where
% multiple graphics applications executing simultaneously on a cloud server
% using a GUI sharing proxy with a non-virtualized GPU (e.g., with PCIe
% pass-through). %\footnote{Note that, even though the GPU is not virtualized, other hardware components, including CPUs, may still be virtualized.}
The system in Figure~\ref{fig:virt_gpu_arch_proxy} operates in the following
steps. When the client's interactive device captures a user input (e.g., a
keystroke, mouse movement or head motion), it sends the input through the
network to a proxy on the cloud server (step \ding{202}), which forwards the
input to the application (step \ding{203}). The proxy is usually a server
application that handles media communication protocols, such as a Virtual
Network Computing (VNC) server with Remote Frame Buffer (RFB) protocol or a
video streaming server with extended Real Time Streaming Protocol
(RTSP)~\cite{1998-Richarson-IC,2013-Huang-MMSys,1998-schulzrinne-RTSP}. After
receiving the input, the application starts frame rendering (step \ding{204}). A
3D application may use a rendering engine that provides functions for drawing
complex objects (e.g., the ``Application 1''), or it may directly call a 2D/3D
library to draw objects from scratch (e.g., ``App 2'').  The rendering engine,
in turn, invokes the 2D/3D library. On Linux, the 2D/3D library is typically
Mesa 3D Graphics Library, which implements the APIs of OpenGL and
Vulkan~\cite{MESA,OpenGL,Vulkan}. Examples of the rendering engine include
Unity, OSVR and OpenVR~\cite{2015-Boger-VR, Unity3D, OpenVR}. To ensure 2D/3D
calls are indeed invoked on the server, 
% and to copy the rendered frames from the GPU,
a graphics interposer library is employed~\cite{2007-Commander-virtualgl}. The
2D/3D library (and the GPU driver) then translates the drawing APIs into GPU
commands to
% into Direct Rendering Manager (DRM) APIs (step 4), which are converted to GPU
% commands to
perform the rendering on the GPU (step \ding{205}). After the frame is rendered
on the GPU, the graphics interposer copies the frame from the GPU (step
\ding{206}-\ding{207}) and push the newly rendered frame to the server proxy
(step \ding{208}). The proxy then compresses and sends the frame over the
network to the client for display (step \ding{209}).

Moreover, as shown in Figure~\ref{fig:virt_gpu_arch_proxy}, multiple 3D
applications can execute simultaneously on the same machine and share hardware
components, such as CPU, memory, GPU and PCIe buses. Each application is
executed in a virtual machine (VM) or a container with virtualized 
GPUs (vGPU)~\cite{2018-nvidia-docker, 2009-Ada-GViM, 2007-Lagar-ICVEE}.

This system architecture has two implications for the benchmarking of cloud
graphics rendering systems. First, as the behaviors of 3D applications are
heavily influenced by user inputs, reliably benchmarking 3D applications
requires generating human-like inputs. Second, cloud graphics rendering system
includes complex and heterogeneous software/hardware components, which must be
properly handled/measured when analyzing performance. In the rest of this paper,
we will describe the design of \CGraph, which overcomes the challenges mention
in Section~\ref{sec:intro} and the above two issues.

%%% Local Variables: 
%%% mode: latex
%%% TeX-master: "paper.tex"
%%% End:
\section{The Design of \CGraph}\label{sec:benchmark_framework}
\begin{figure*}
  \centering
    \includegraphics[width=15cm]{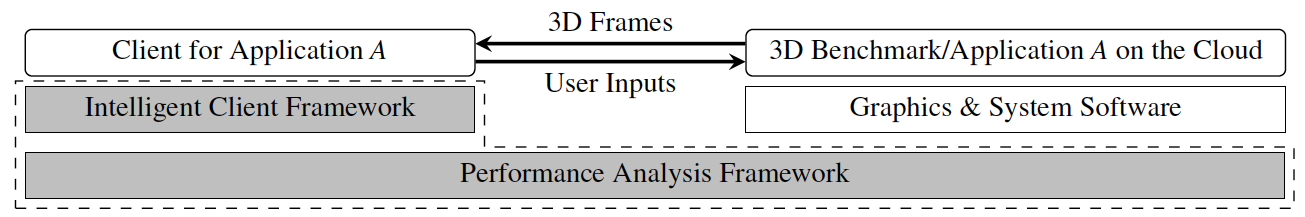}
  \vspace{-2mm}
  \caption{Overview of the \CGraph benchmarking framework (surrounded by the dashed box).}
  \label{fig:framework_overview}
  \vspace{-3mm}
\end{figure*}

Figure~\ref{fig:framework_overview} shows the components of \CGraph benchmarking
framework. A main component of \CGraph is the intelligent client framework that
is used to generate clients with human-like actions to interact with 3D
applications. %Inspired by autonomous driving, we proposed to design the intelligent client framework with Computer Vision (CV) to recognize objects in 3D frames and with RNN to learn how to interact with the application like a human~\cite{2012-Alex-NIPS,2011-Socher-ICML,1997-Sperduti-NN}. %Besides accurately simulating human actions, another benefit of using CV/RNN is that a new intelligent client can be easily built for a new 3D application without knowing it internal designs.
%Besides the intelligent client,
The other main component of \CGraph is the performance analysis framework, which
spans over the client and the server to provide reliable performance
measurements. The rest of this section describes these two components in detail.

\subsection{Intelligent Client Framework Design}\label{sec:intelligent_client}
\begin{figure*}
  \centering
   \includegraphics[width=15cm]{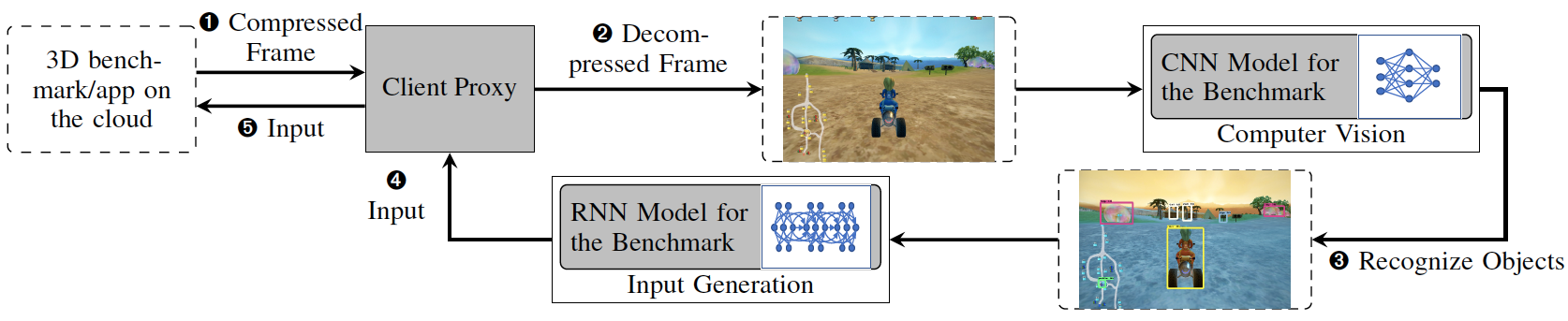}
  \vspace{-2mm}
  \caption{Overview of the intelligent client. The image is obtained from a racing game, SuperTuxKart~\cite{SuperTuxKart}.}
  \label{fig:client_framework}
\end{figure*}

\textbf{Overview} The intelligent client framework allows building an
intelligent client for a 3D application by learning how to properly interact
with this application from recorded human actions. More specifically, for a 3D
application, an RNN model is trained based on recorded human actions under a
scene of this application~\cite{2011-Socher-ICML}. To improve the RNN model's
accuracy, the objects in the frames are first recognized using CV with
Convolutional Neural Network (CNN)~\cite{2012-Alex-NIPS}.

Figure~\ref{fig:client_framework} gives an overview of a client obtained with
the intelligent client framework, which operates in the following steps. After a
compressed frame is sent over the network to the intelligent client (step
\ding{202}), it is first decompressed (step
\ding{203}). %. %This proxy is mainly responsible to communicate with the server proxy (Figure~\ref{fig:non_virt_gpu_arch}) or the X-Server (Figure~\ref{fig:virt_gpu_arch}).
%If the image is compressed, the proxy is also responsible to 
The decompressed frame is then processed by a CNN model to recognize its objects
(step \ding{204}).
% This CV program has at least two ways to recognize objects: using template matching or Convolutional Neural Network (CNN).
The types and coordinates of the recognized objects are then sent to an
RNN model to generate user inputs that mimic real human actions (step
\ding{205}).
%Note that, for some 3D applications, instead of RNN, simple
%rule-based input generators may suffice.
These inputs are eventually sent back to the client proxy, which encodes these
actions into network packages and sends them to the benchmark (step \ding{206}).
With CNN and RNN models, the clients can properly interact with 3D applications
with random frames and under random networking/system latency. By generating actions
purely based on frames, the clients can be built for 3D applications without
knowing their internal designs or modifying these
applications. %Currently developments in AI also allows the CNN and RNN models to have fast inferences that can match the speed of human reactions~\cite{2012-Pulli-Queue}.

\textbf{Model Training} Each 3D application/benchmark has its own CNN/RNN
models, which are trained from a recorded session of human actions under an
application scene. The intelligent client framework provides tools to perform
this recording. %using image and user-input capturing.
Each recorded session includes a sequence of frames and the corresponding human
actions to each frame. To train a CNN model, the objects in the frames need to
be manually labeled. The labeled frames are then fed into a machine-learning
(ML) package to train the CNN model. The manual frame labeling is generally fast
and takes about 4 hours for one 3D application in our experience, as only the
objects that can determine the user inputs need to be labeled.

An RNN model can also be trained using the recorded session. The recorded frames
are first processed by its CNN model to recognize the objects. After the
recognition, the recorded data are converted into a training data set where the
features are the objects in a frame and the labels are the corresponding human
actions. An RNN model can then be trained to learn how to respond to the objects
in a frame like a real human. Note that, our goal is not to train an AI to
compete with human. Instead, we aim at training an RNN model to mimic human
actions under varying system latency and frame randomness, so that the
performance results obtained with the RNN-generated actions are similar to those
obtained with real human users. Because a trained RNN model is executed on the
same scene where it is trained, the model is likely to work well as long as it
has low training loss.

\textbf{Implementation} We implemented the training and inference of the CNN and
RNN models with Tensorflow~\cite{2016-Abadi-OSDI}. The actual CNN model used is
MobileNets~\cite{2017-hHward-MobileNets}. The actual RNN model used is Long
Short-Term Memory (LSTM)~\cite{1997-Long-NC}.
%%% Local Variables: 
%%% mode: latex
%%% TeX-master: "paper.tex"
%%% End:
\subsection{Performance Analysis Framework Design}\label{sec:perf_framework}
\textbf{Overview} The performance analysis framework provides performance
measurements of 3D applications and cloud graphics rendering
systems. Performance measurements include frame rate (FPS, frames-per-second),
the latencies of each stage involved in the handling of a user input and the
delivery its response frame, as well as system-level and architecture-level
resource usages. As stated in Section~\ref{sec:intro}, designing this framework
has two difficulties. The first difficulty is to accurately track and associate
the processing of an user input and the rendering of its response frame. The
second difficulty is to measure the performance of the complex and heterogeneous
software/hardware components. This section describes how \CGraph overcomes these
difficulties.
% of the diverse types of hardware and software components used in the graphics
% rendering and network communication.

\begin{figure*}
  \centering
  \includegraphics[width=15cm]{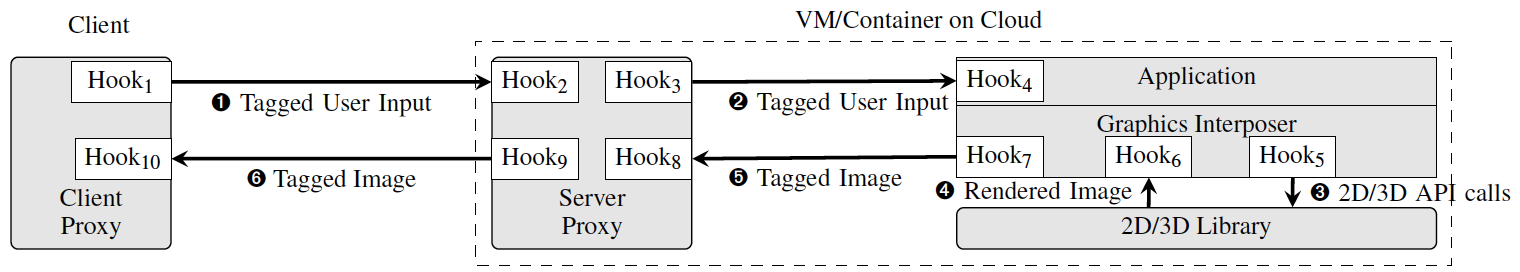}
  \vspace{-2mm}
  \caption{Using API hooks to track the processing of a user input and the rendering of its reponse frame.}
  \label{fig:api_hooks}
\end{figure*}

\textbf{Tracking User Input Processing}. %Accurately measuring the round-trip time (RTT) and the time spent in each software/hardware component of the rendering of a response frame to a user input is crucial to the performance analysis of cloud graphics rendering. However, the complex graphics software/hardware stack, the parallelism in the rendering process and the fact that rendering happens without user input (3D application automatically updates its frames), make tracking the processing and rendering of user inputs very difficult~\cite{2003-Nieh-TOCS}.
%requires correctly determining which image is  the response to a user input. This determination is particularly challenging as GUI applications constantly refresh their UIs even when there is no input. Hence, a GUI application may automatically generate many UI images after it receives a user input and before it produces the response image, making it very difficult to know which image is actually the response. This determination is further complicated by the parallel processing of the user inputs and response image rendering, and by the need to send the input and image across of the boundaries of many processes, hardware and the network.
To track the input processing, we tag the input from the client and use the tag
to identify every stage of the input processing. More specifically, at the
beginning of each processing stage, the tag of the corresponding user input is
extracted from the input data. At the end of the stage, the tag is added to the
output data, allowing the next stage to extract it. For graphics rendering, the
begin and end of each stage can be determined based on the invocation of
specific OpenGL and X-Window APIs, and the tags can be passed along as
input/output data to these APIs. The invocations to these APIs can be
intercepted with API hooks, allowing extracting/adding the tags in these hooks.

\begin{table}
  \fontsize{9}{10}\selectfont
  \centering
  \begin{tabular}{|c|l|}
    \hline
    Hooks & Intercepted APIs  \\ \hline
    Hook$_4$ & XNextEvent, glutKeyboardFunc \\ \hline
    Hook$_5$ & glxSwapBuffer, glutSwapBuffers\\ \hline
    Hook$_6$ & glReadBuffer, glReadPixel\\ \hline
    Hook$_7$ & XShmPutImage, glMapBuffer\\ \hline
  \end{tabular}
  \vspace{-2mm}
  \caption{Some of the APIs intercepted at the API hooks.}
  \label{tab:hookable_apis}
  \vspace{-4mm}
\end{table}

Figure~\ref{fig:api_hooks} illustrates the API-hook-based input-tracking
technique. For now, we assume a sequential graphics rendering process. To track
an input, hook$_1$ at the client proxy gives every input a unique tag and sends
the tag with the input to the server proxy. Upon receiving the input, hook$_2$
at the server proxy extracts the tag from the network package. The tag is then
forwarded to the application with its input by hook$_3$. When the application
receives the input, the tag is extracted at hook$_4$ and saved. Hook$_5$ marks
the start of the GPU rendering, there is no need to send the tag to GPU. At
hook$_6$, the saved tag is embedded into the pixels of the rendered frame (the
old pixels are stored in shared memory). Embedding the tag in pixels ensures
that the tag survives the inter-process communications between the application
and server proxy. After the server proxy receives the tagged frame at hook$_8$,
it extracts the tag, restores the modify pixels and sends the frame with the tag
to the client. Once hook$_{10}$ at the client proxy receives the tagged frame,
it matches the tag with a previously sent user input, which finishes the
tracking.  Table~\ref{tab:hookable_apis} gives some examples of the APIs that
can be intercepted from hook$_4$ to hook$_7$. The other hooks in the server and
client proxies can be easily identified using their source code.

\begin{figure*}
  \centering
  \includegraphics[width=15cm]{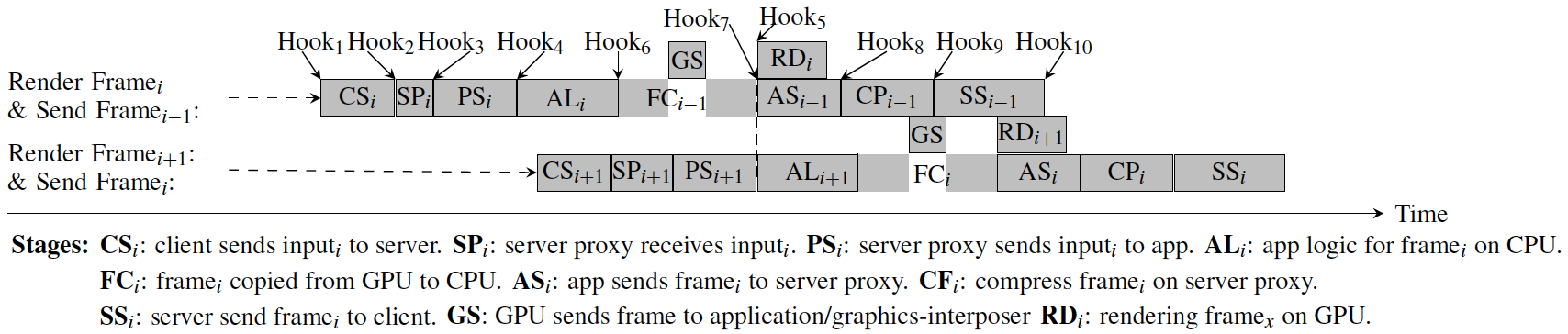}
  \vspace{-2mm}
  \caption{A typical software pipeline for open-source remote 3D-graphics rendering on Linux.}
  \label{fig:graphics_pipeline}
\end{figure*}

However, instead of the above sequential rendering process, modern graphics
applications typical employ %multiple display buffers and
software pipelines to parallelize the rendering for better
performance. \FIG~\ref{fig:graphics_pipeline} shows the typical stages of this
pipeline for remote 3D-graphics rendering when rendering two frames, frame$_i$
and frame$_{i+1}$.
%The rendering of two frames
%(frame$_i$ to frame$_{i+1}$) is shown in \FIG~\ref{fig:graphics_pipeline}, with
%the second frame being the application's automatic frame update.
As
\FIG~\ref{fig:graphics_pipeline} shows, in each pass of the pipeline, a new
frame is rendered, and the previous frame is copied and sent to the clients. For
example, in the first row of \FIG~\ref{fig:graphics_pipeline}, frame$_i$ is
rendered based on input$_i$, while frame$_{i-1}$ is copied from the GPU and sent
to the client.

Note that, \FIG~\ref{fig:graphics_pipeline} shows the pipeline for the cloud
rendering system analyzed in Section~\ref{sec:perf_analysis}.  In this system,
the stages of application-logic (AL) and frame-copy (FC) are carried out by the
same thread due to the difficulty to know when a frame is completely rendered in
the GPU. Therefore, AL and FC stages cannot overlap, and the next AL stage must
start after the previous FC stage is finished. Nonetheless, any other two stages
in this pipeline can overlap, as they are not carried out by the same
thread/processor.

The main benefit of this software pipelining is that it allows the CPU and GPU
to execute simultaneously. For example, as shown in
\FIG~\ref{fig:graphics_pipeline}, when frame$_i$ is being rendered on the GPU
(stage RD$_i$), the CPU is working on the application logic for frame$_{i+1}$
(stage AL$_{i+1}$) and sending frame$_{i-1}$ (stage AS$_{i-1}$) using two
threads/cores.  The tag-based input tracking still works for this parallel
rendering, as long as the tracking implementation is aware that the
processing/rendering of an input spans over two passes of the pipeline.

\textbf{Performance Measurements for Diverse Components}.
% Various techniques were employed by \CGraph to handle the diverse types of
% components.
The API hooks also allow measuring the execution times (latencies) of each stage
involved in the rendering. A hook records a timestamp when it intercepts an API
call. The differences between the timestamps of two hooks then give the time
spent in each stage. %Table~\ref{tab:measured_times} lists some of the execution times/latencies that can be measured.
For example, the time difference between the hook$_{10}$ and hook$_1$ with
matching tag gives the round-trip time (RTT) to handle a user
input.% and the time difference between Hook$_{6}$ and Hook$_7$ gives the frame copying time.

However, the time measured with the hook's CPU timestamps cannot give GPU
processing time.  To obtain GPU time, we use the time-querying functionality of
OpenGL~\cite{OpenGL-TimeQuery}. Start and stop querying statements are inserted
into the hooks to measure the GPU time spent in each stage. For example, the time
query starts at a hook$_{5}$ and ends at the subsequently-invoked hook$_6$ gives
the GPU time to render a
frame.

\CGraph also measures FPS and resource usages. The FPS is obtained by counting
the frames at the server and client proxies. System-level resource usages, such
as CPU/GPU and memory utilizations, are obtained from the OS and GPU
drivers~\cite{AMD-ROCM-SMI,NVIDIA-NVML}. Architecture-level resource utilization
is measured using hardware performance monitoring units
(PMU). %These PMUs provide detailed information about the utilization of many hardware components, such as the cache misses on the CPU and GPU.
CPU PMU readings for each stage are obtained by using PAPI inside the API
hooks~\cite{2010-Terpstra-THPC}.  The PMUs on AMD GPUs are queried using AMD's
GPU Performance API~\cite{AMD-GPA}. For NVidia GPU, an external tool, NSight
Graphics, is used to read PMUs, as NVidia does not support programmable PMUs
reading for graphics rendering on Linux.

\textbf{Performance Measurement Extensibility and Overhead}. One benefit of
using API hooks is that it does not require modifying 3D applications. Our
performance analysis framework can be applied to any proprietary 3D
applications, as long as these applications invoke standard 3D APIs, such as
those given in \TAB~\ref{tab:hookable_apis}. As later shown in the experimental
evaluation (Section~\ref{sec:evaluation}), these API hooks also incur little
overhead. However, the time queries used to measure the GPU performance may
stall the CPU and thus incur a high overhead. To mitigate the impact of these
stalls, we used two query buffers and switched them between frames.

%%% Local Variables: 
%%% mode: latex
%%% TeX-master: "paper.tex"
%%% End:
\subsection{The Benchmark Suite}\label{sec:benchmark_suite}
\begin{table}[]
  \fontsize{8}{9}\selectfont
  \centering
  \begin{tabular}{|l|l|l}
    \hline
    Application Area      & Benchmark  \\ \hline
    Game: Racing      & SuperTuxKart (STK)~\cite{SuperTuxKart} \\ \hline
    Game: Real-time Strategy    & 0 A.D. (0AD)~\cite{Game0AD} \\ \hline
    Game: First-person Shoot & Red Eclipse (RE)~\cite{RedEclipse}  \\ \hline
    Game: Online Battle Arena & DoTA2 (D2)~\cite{DOTA2} \\\hline
    VR: Education/Game & InMind (IM)~\cite{InMind} \\ \hline
    VR: Heatlh     & IMHOTEP (ITP)~\cite{2018-Pfeiffer-IJCARS}      \\ \hline
  \end{tabular}
  \caption{Applications included in our benchmark suite.}
  \label{tab:benchmark_list}
  \vspace{-4mm}
\end{table}

With \CGraph, we designed a benchmark suite, which contains four computer games
and two VR applications. All benchmarks are from real applications and cover
popular game genres and usage cases. Table~\ref{tab:benchmark_list} lists these
benchmarks. Among the six benchmarks, \textit{Dota2} and \textit{InMind} are
closed-source. Note that, as \CGraph is designed to be extensible, new 3D
applications can be easily added in the future.

%%% Local Variables: 
%%% mode: latex
%%% TeX-master: "paper.tex"
%%% End:

%%% Local Variables: 
%%% mode: latex
%%% TeX-master: "paper.tex"
%%% End:
\section{Evaluation}\label{sec:evaluation}
This section provides the experimental evaluation of the reliability/accuracy and overhead of \CGraph.

\begin{figure}
  \centering
  \includegraphics[width=8cm]{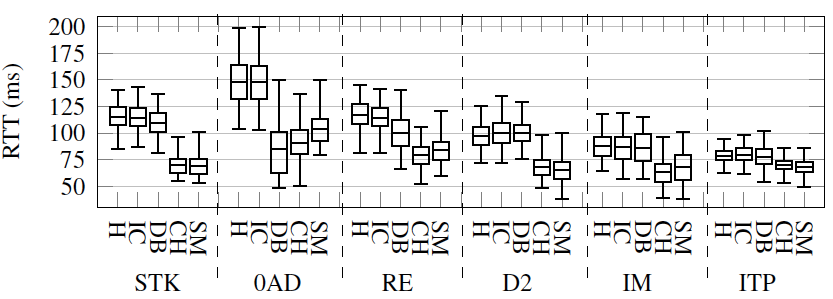}
  \vspace{-4mm}
  \caption{The performance (RTT) distributions obtained with human users (H),
    \CGraph's intelligent clients (IC), DeskBench~\cite{2005-Zeldovich-ATC} (DB),
    Chen et al.~\cite{2014-Chen-TOM} (CH), and Slow-Motion~\cite{2003-Nieh-TOCS}
    (SM).}
  \label{fig:human_vs_bot}
\end{figure}

\begin{table}
  \fontsize{8}{9}\selectfont
\begin{tabular}{|l|c|c|c|c|c|c|c|}
  \hline
  & STK    & 0AD    & RE     & D2     & IM     & ITP    & Avg    \\ \hline
  \CGraph  & 0.8\%   & 0.1\%    & 2.5\%   & 3.2\% & 1.3\% & 2.0\% & 1.6\% \\ \hline
  DB        & 5.4\%   & 42.9\%  & 14.6\% & 3.3\% & 1.3\% & 2.3\% & 11.6\% \\ \hline
  CH   & 39.5\%  & 38.9\%  & 32.6\% & 29.9\%  & 11.4\%  & 27.8\% & 30.0\% \\ \hline
  SM   & 39.8\%  & 30.1\%  & 28.2\% & 32.7\%  & 13.7\%  & 22.7\%  & 27.9\%  \\ \hline
\end{tabular}
\caption{Percentage errors for the means of the RTTs obtained with \CGraph's IC,
  DeskBench~\cite{2005-Zeldovich-ATC} (DB), Chen et al.~\cite{2014-Chen-TOM}
  (CH), and Slow-Motion~\cite{2003-Nieh-TOCS} (SM), when compared to the mean
  RTTs obtained with human users.}
\label{tab:mean_RTT_erros}
\end{table}

\begin{figure}
  \centering
  \includegraphics[width=8cm]{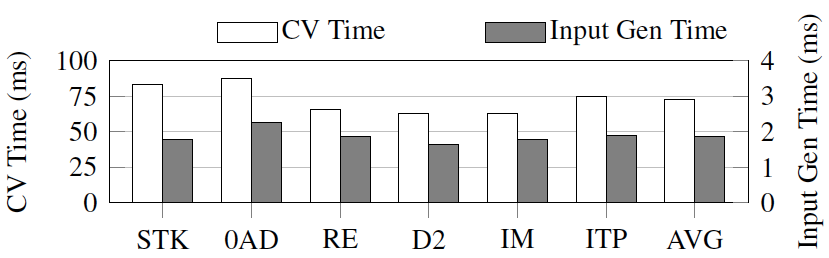}
  \vspace{-4mm}
  \caption{Computer vision and input generation time.}
  \label{fig:client_perf}
  \vspace{-4mm}
\end{figure}

\textbf{Experiment Setup} The benchmarks were executed on a server with an
8-core Intel i7-7820x CPU, 16GB memory and an NVIDIA GTX1080Ti GPU with 11GB GPU
memory. The clients consisted of four machines each with a 4-core Intel i5-7400
CPU and 8GB memory. The server and clients were connected using 1Gbps
networks. 1Gbps network was chosen because it behaved similarly to 5G cellular
network in terms of the frame-transmitting latency as shown later in
Section~\ref{sec:single_bench_app_perf}. Precision Time
Protocol~\cite{2002-Eison-PTP} was used to synchronize the time between the
server and clients.

The server and clients run Ubuntu 16.04 as the OS and TurboVNC
2.1.90~\cite{TurboVNC} as the rendering system. We chose VNC as it has complete
support for 3D rendering. The other open-source solution,
GamingAnywhere~\cite{2013-Huang-MMSys}, failed to run all ofour benchmarks. To
the best of our knowledge, all VNC implementations (and even the non-VNC
proprietary NX technology~\cite{2003-Pinzari-NX}) required TurboVNC's graphics
interposer, VirtualGL~\cite{2007-Commander-virtualgl}, to support 3D
rendering. Therefore, we evaluated TurboVNC, as it represents the
state-of-the-art open-source remote 3D rendering. We modified TurboVNC to
support VR device inputs.  All benchmarks were executed at a resolution of
1920$\times$1080 with maximized visual effects.

\textbf{Intelligent Client Accuracy Evaluation.} To evaluate if the intelligent
clients (ICs) indeed allow reliable and accurate performance results, we
compared the benchmarks' behaviors under the ICs and human interactions. More
specifically, each benchmark was executed using its IC and was also played/used
by a real human user for three 15-minute sessions each (results were stable
after 10 min). We then compared the performance results obtained from the two
types of executions, including the latency, FPS, and CPU/GPU utilization.
\FIG~\ref{fig:human_vs_bot} shows the round-trip time (RTT) that it took to
process input for each benchmark when executed with the IC and the human
user. For each execution, \FIG~\ref{fig:human_vs_bot} shows the mean, 1\%-tile,
25\%-tile, 75\%-tile and 99\%-tile of the measured RTTs.  As
\FIG~\ref{fig:human_vs_bot} shows, the RTTs obtained with IC were very similar
to those from the human. \TAB~\ref{tab:mean_RTT_erros} also gives the percentage
errors of the means of the RTTs obtained with our IC.  The maximum percentage
error for the mean-RTT for IC is only 2.5\%, and average error for IC is only
1.6\%,
% Moreover, we employed multinomial likelihood~\cite{2007-Shlens-SNL} to compare
% the performance distributions obtained with the ICs and the human
% user. Intuitively, given an observed distributions $d_o$ and a model distribution
% $d$, the multinomial likelihood gives the probability that $d_o$ is indeed an
% observation from $d$~\cite{2007-Shlens-SNL}. For the RTT distributions
% comparison, multinomial likelihood gives the probability that a IC distribution
% is
% For the rest of this paper, we simply refers to the multinomial likelihood as
% the \textit{accuracy} of an RTT distribution obtained with \CGraph and other
% performance measuring techniques.
the RTTs from the IC and human runs were also similar. The data for other
performance metrics were also similar for both runs. However, limited by space,
other performance metric results are omitted.

% In fact, random and trace-based inputs could be event worse, due to mismatching
% inputs and game scences, which would lose game control flow because of
% randomness and dynamic network latency. Event through we tried to avoid these
% negative influence and get the best possible results of random inputs and
% trace-based inputs, ICs still behave better than these two. These results show
% that the ICs can indeed accurately mimic real human actions.

\textbf{Intelligent Client Speed Evaluation.} Figure~\ref{fig:client_perf} gives
the average times that it took to conduct CV (CNN) and generate input (RNN) for
each benchmark. As the figure shows, the clients have fast inference times, with
an overall average of 72.7ms for CV and 1.9ms for input generations. This fast
inference allows the ICs to generate 804 actions per minute (APM) on average,
which is faster than professional game players (about 300
APM)~\cite{2008-McCoy-AAAI}, showing that the ICs can generate inputs fast
enough to mimic human reaction speed.
% The longest CV time is 87.5ms for \textit{0AD}, which requires recognizes
% several complex game units. With the action generation time, the IC for 0AD
% has an APM of 668, which is still faster enough to mimic human actions. For
% \textit{RedEclipse}, its inputs are generated using rules instead of
% RNN. Therefore, its input generation time is only 2.21ms.

\textbf{\CGraph Overhead Evaluation.}  To evaluate the overhead of \CGraph, we
executed each benchmark with and without the performance analysis framework. For
the run without the performance analysis framework, native TurboVNC is used with
our ICs. As the native TurboVNC does not provide RTT readings, we compared the
FPS of both runs.
% Figure~\ref{fig:perf_frmk_overhead} gives the overhead (percentage differences
% in FPSes) of the performance analysis framework of
% \CGraph. As Figure~\ref{fig:perf_frmk_overhead} shows,
Our results show the performance analysis framework has low overhead. The FPS
reduction was only 2.7\% on average (5\% at maximum) for all benchmarks. This
low overhead is partially due to our use of double-buffers when querying GPU
execution times. Without these double-buffers, the overhead was up to 10\%.

\textbf{Comparison with Prior Work.} To show the importance of properly handling
irregular/random objects and tracking inputs, we also compared \CGraph with
three prior performance measuring techniques for VDI and cloud gaming.

We first compared \CGraph with DeskBench~\cite{2009-Rhee-INM}. DeskBench was
based on VNCPlay~\cite{2005-Zeldovich-ATC} and replayed recorded human actions
to generate inputs. However, DeskBench did not only record an action, it also
recorded the screen frame when this action was issued. During replay, the action
was only issued when the displayed frame was similar to the recorded frame. With
this frame comparison, DeskBench (and VNCPaly) only issued an action when the
expected object was displayed, and thus, was capable to handle network latency
variation. Note that, the ``similarity'' between frames was a tune-able
parameter for DeskBench. We tested with several parameter values following the
methodology presented by DeskBench and reported the DeskBench's results using
the best parameter we found. Additionally, as DeskBench did not provide input
tracking, it was only used to generate inputs, and \CGraph's performance
framework was used to collect performance data. \FIG~\ref{fig:human_vs_bot} and
\TAB~\ref{tab:mean_RTT_erros} also give the RTT distributions and errors
obtained with DeskBench. The average error of the mean-RTT obtained with
DeskBench was 11.64\%, which was considerably higher than the 1.6\% error of
\CGraph. DeskBench was designed for 2D applications with well-shaped and placed
objects (e.g., icons and texts), where simply comparing pixels can determine if
an object is shown or not. However, for 3D games, even the same object can have
different pixels and locations depending on the viewing angle and the flow of
events. Hence, simply comparing the pixels is practically impossible to
determine the existence of an object, causing DeskBench to frequently delay an
action.

We also compared \CGraph with a cloud gaming performance analysis methodology
presented by Chen et al.~\cite{2014-Chen-TOM}. In this methodology, the authors
generated inputs with human players. This methodology did not provide input
tracking, and hence, could not measure RTT at the client. Therefore, it had to
compute the RTT by summing the time of the stages of CS, SP, AL, CP, and SS of
the software pipeline. \FIG~\ref{fig:human_vs_bot} and
\TAB~\ref{tab:mean_RTT_erros} also give the RTT distributions and errors
obtained with this methodology. The average error of the mean-RTT obtained with
this methodology was 30.0\%, which was also much higher than the 1.6\% error of
\CGraph. There are two issues with this methodology because of the lack of input
tracking. First, the AL latency in this methodology was obtained offline without
the VNC server proxy. This offline measurement gave lower AL latency than that
obtained during online execution, because it eliminated the resource contention
between the game and the VNC server proxy. Second, with input tracking, the
methodology could not measure the delays of the inter-process communication
stages, including PS, FC, and AS. Because of these two issues, Chen et al.'s
methodology usually reported smaller RTTs than those directly measured at the
client.

The last comparison was conducted with a VDI performance measuring technique
call Slow-Motion~\cite{2003-Nieh-TOCS,2010-Berryman-CloudCom}. Slow-Motion was
designed to determine the RTT of one frame. Slow-Motion injected delays into the
cloud rendering system to only allow one input/frame being processed at a time
-- only after an input was processed, its frame was rendered and sent to the
client, could the processing of the next input/frame start. %In the extreme case,
% the system might be reduced to 1FPS for performance measurement.
By allowing only one frame at a time, it was trivial to associate an input with
its response frame. Note that, as Slow-Motion did not include an input
generation technique, \CGraph's IC was used to generate the
inputs. \FIG~\ref{fig:human_vs_bot} also shows the RTT distributions obtained
with Slow-Motion. The average error of the mean-RTT obtained with this
methodology was 27.9\%, which was also higher than the 1.6\% error of
\CGraph. The main issue of Slow-Motion was that the injected delay changed the
resource usage and the behavior of the benchmark and the VNC server proxy (which
was also noted by the original authors~\cite{2003-Nieh-TOCS}). Because only one
frame was rendered at a time, the resource contention caused by parallel
processing/rendering of the inputs and frames was eliminated, and the resource
contention between the benchmark and the server proxy was also
reduced. Consequently, Slow-Motion typically reported smaller RTTs than those
observed with a system executing at full capacity.

%Need update: 
% Furthermore, to show the importance of having intelligence in the clients,
% we also compared the performance results obtained from three other works (VNCPlay, chen, and slow motion) that are also shown in
% \FIG~\ref{fig:human_vs_bot}. The maximum errors of mean-RTTs for VNCPlay (a record and replay scheme)
% is 42.93\% (\textit{0AD}) and the average error is 11.59\%. The maximum and
% average errors of mean-RTTs for Chen's work are 39.47\% (\textit{STK}) and
% 30.0\% respectively. Similar high errors were also observed for FPS and CPU/GPU
% utilization. These high errors were mainly because random inputs and recorded
% traces caused the benchmarks to stuck (fail to proceed). These high errors show
% that neither recorded traces nor random inputs can be used to obtain reliable
% performance measurements.

%%% Local Variables: 
%%% mode: latex
%%% TeX-master: "paper.tex"
%%% End:

\vspace{-3mm}
\section{Performance Analysis of Cloud Rendering System and 3D Applications}\label{sec:perf_analysis}

\subsection{Perf. Analysis with A Single Benchmark}\label{sec:single_benchmark}
This section provides the performance analysis results with a single benchmark, which was executed using the same methodology given in Section~\ref{sec:evaluation}.

\begin{figure}
  \centering
  \includegraphics[width=8cm]{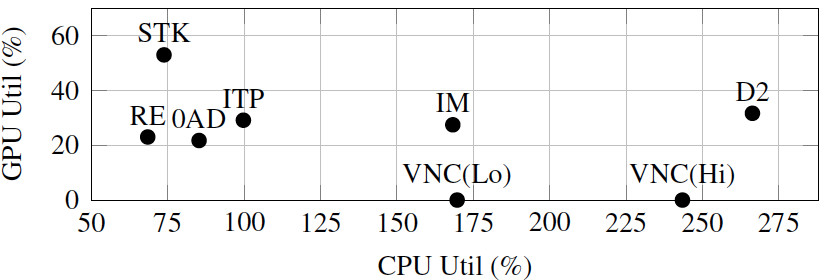}
  \vspace{-4mm}
  \caption{CPU and GPU utilization for each benchmark.}
  \label{fig:cpu_gpu_util}
\end{figure}

% \begin{figure}
%   \centering
%   \input{figures/single_bench_memory_usage}
%   \caption{CPU and GPU memory usages for each benchmark.}
%   \label{fig:single_bench_mem_usage}
% \end{figure}

\begin{figure}
  \centering
  \includegraphics[width=8cm]{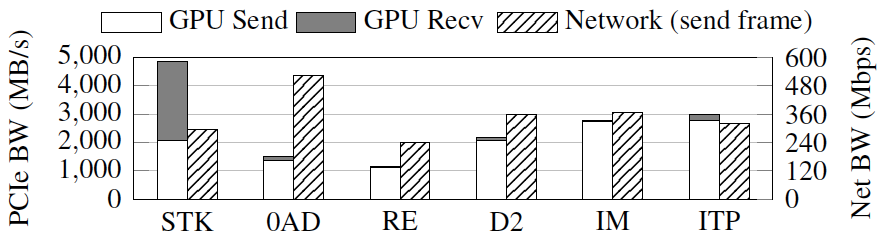}
  \vspace{-4mm}
  \caption{Network and PCIe (send-to and received-at the GPU) bandwidth usages for each benchmark.}
  \label{fig:single_bench_net_pcie_usage}
\end{figure}

\subsubsection{System-level Resource
  Utilization}\label{sec:single_bench_system_util}
\FIG~\ref{fig:cpu_gpu_util} gives the CPU and GPU utilization of each
benchmark. The CPU utilization of these benchmarks had a high variation, ranging
from 68\% (\textit{RedEclipse}) to 266\% (\textit{Dota2}).
% \textit{Dota2}'s high CPU utilization is likely due to its strategy-game
% nature and its sophisticated AI.  Most of the benchmarks had less than 35\%
% GPU utilization mainly because a high-end GPU was used in the
% experiments. %\textit{SuperTuxKart} had the highest
% GPU utilization of 53\%, as it rendered an underwater race track.
The GPU utilization also had a high variation, ranging from 22\% to 53\%.  The VNC
server also had considerable CPU utilization, which varied from 169\% to 243\%,
depending on the FPS and frame compression difficulty.
% We observed that the majority of VNC server's CPU time was spent on image
% compression.
%\FIG~\ref{fig:single_bench_mem_usage} gives the memory usages of each
%benchmark.
The CPU memory usages also vary considerably, ranging from 600MB (\textit{Dota2}) to
nearly 4GB (\textit{InMind}). The GPU memory usages of these benchmarks were
less than 800MB, which is similar to the 1GB-2GB GPU memory requirements of
recent popular games.

\FIG~\ref{fig:single_bench_net_pcie_usage} shows the network and PCIe bandwidth
usages for each benchmark. For network usage, only the bandwidth usage of
sending the frames to the client is shown, as sending the inputs from the
clients used only 1.5Mpbs. The network usages of these benchmarks were below
600Mpbs, which is lower than the maximum bandwidth of the coming 5G cellular
network and 10Gpbs broadband. Similarly, all benchmarks used less than 5GB/s on
the PCIe bus, which is well below the 31.5GB/s maximum bandwidth
of %(31.5GB/s bidirectional) of
PCIe3. Except for \textit{SuperTuxKart}, all benchmarks sent limited amount of
data from the CPU to GPU, suggesting most of their rendering data were stored on
the GPU. The exception of \textit{SuperTuxKart} was likely due to its frequent
and drastic changes in the rendered frames.  For all benchmarks, there is high
PCIe bandwidth usage from GPU to CPU, which represented the data used for copying
rendered frames from GPU to CPU.
%the main use of
%PCIe bus was to copy rendered frames from GPU to CPU.

% \begin{figure}
%   \centering
%   \input{figures/single_bench_rtt}
%   \caption{Breakdown of RTTs for each benchmark.}
%   \label{fig:single_bench_rtt}
% \end{figure}

% \begin{figure}
%   \centering
%   \input{figures/single_bench_server_time}
%   \caption{Breakdown of server latencies for each benchmark.}
%   \label{fig:single_bench_server_time}
% \end{figure}

% \begin{figure}
%   \centering
%   \input{figures/single_bench_app_time}
%   \caption{Breakdown of application latencies for each benchmark.}
%   \label{fig:single_bench_app_time}
% \end{figure}

\begin{figure}
  \centering
  \includegraphics[width=8cm]{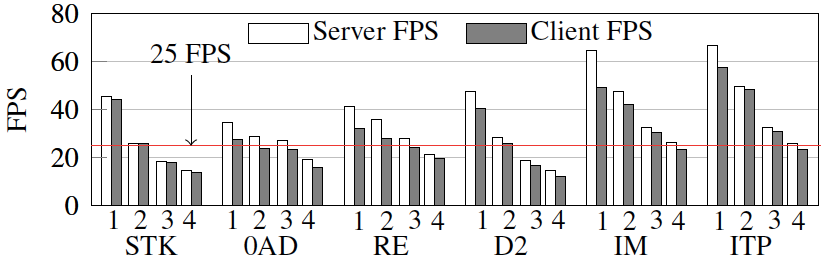}
  \vspace{-4mm}
  \caption{Server and client FPS when executing one to four instances of the same benchmark on the server.}
  \label{fig:multi_bench_fps}
\end{figure}

\subsubsection{Application Performance}\label{sec:single_bench_app_perf}
\FIG~\ref{fig:multi_bench_fps} gives the server and client FPS for each
benchmark when one to four instances of the same benchmark was executed on the
server. Here, we focus on the FPS for one instance (i.e., the bars with x-axis
label ``1''). Server FPS measured the number of frames that were generated at
the server in one second.  Client FPS measured the number of frames the client
received in one second.  The lowest client FPS was 27 (for \textit{0AD}), which
is still higher than the minimum 25 FPS quality-of-service (QoS) requirement for
3D applications, showing the feasibility of cloud graphics
rendering~\cite{2004-Rumsey-AES}.

\begin{figure}
  \centering
  \includegraphics[width=8cm]{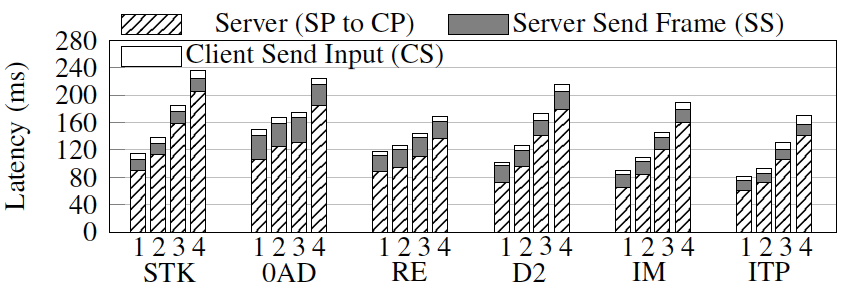}
  \vspace{-4mm}
  \caption{RTT breakdown when executing one to four instances of the same benchmark on the server.}
  \label{fig:multi_bench_rtt}
\end{figure}

\FIG~\ref{fig:multi_bench_rtt} gives the average RTTs of handling an input for
each benchmark. Again, we focus on the RTTs for one instance (i.e., the bars
with x-axis label ``1'').  These RTTs are broken down into the time the sever
spent on handling the input and the network times for sending the inputs and
frames.  For all benchmarks, the network latency for sending inputs (stage CS)
was very small (< 10ms). The network latency for sending frames (stage SS)
ranged from 14ms to 35ms, which was
% , with most benchmarks experienced a frame-sending latency less than
% 20ms. These latencies
similar to those reported by prior work with 4G/5G cellular network, suggesting
our 1Gpbs network is close to the real use
case~\cite{2018-Tan-SIGMETRICS}. %\textit{0AD} had the highest frame-sending
%latency due to its low frame compression ratio, resulting in large frames sent
%to the client.
The largest component of RTT was always the time that the server took to process
inputs, which include all stages from SP to CP.  This server processing time
ranged from 61ms to 106ms. Such high server time indicates that cloud system
design is crucial to ensuring good performance.

\begin{figure}
  \centering
  \includegraphics[width=8cm]{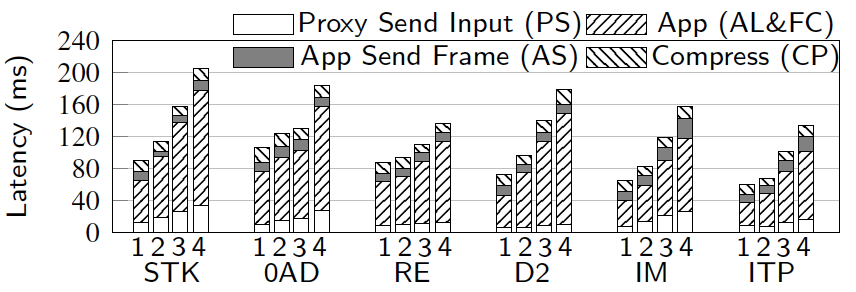}
  \vspace{-4mm}
  \caption{Server time breakdown when executing one to four instances of the same benchmark on the server.}
  \label{fig:multi_bench_server_time}
\end{figure}

In \FIG~\ref{fig:multi_bench_server_time}, the server time is further broken
down into the time of VNC sending inputs to the benchmark (stage PS), the
application execution time (stage AL, FC, and RD), benchmark sending frame to
VNC (stage AS) and the time of VNC compressing frames (stage CP). Note that, the
time for stage SP is omitted because it was too small ($< 1m$) to be visible in
the figure.
% For all benchmarks, the time for VNC-sending-input stage was
% consistently around 10ms. The time for frame-sending stage was also very small
% ($\le$ 11ms). % due to use the shared memory in TurboVNC.
% Compression time ranged from 12ms to 18ms, depending on the complexity of the
% frames.
As \FIG~\ref{fig:multi_bench_server_time} shows, for all benchmarks, the main
component of the server processing time is the application execution time, where
the execution times for other stages (i.e, PS, AS and CP) were less than 18ms.

%the times for PS, AS and CP stages were less than 18ms
%which were much smaller than the application execution time.
% The largest component of server time was always the benchmark execution time,

\begin{figure}
  \centering
  \includegraphics[width=8cm]{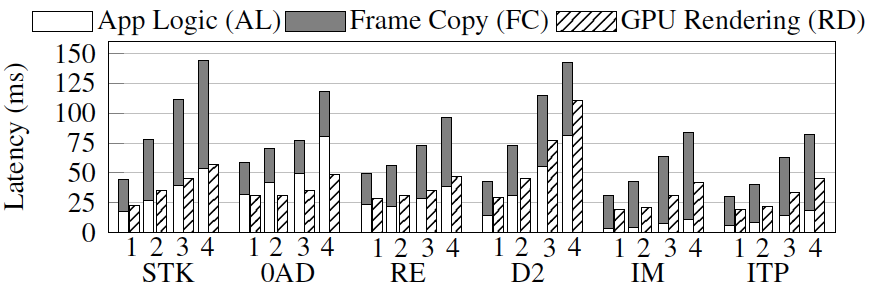}
  \vspace{-4mm}
  \caption{Application time breakdown when executing one to four instances of a
    benchmark on a server.}
  \label{fig:multi_bench_app_time}
\end{figure}

The application execution time was further broken down in
\FIG~\ref{fig:multi_bench_app_time}.
% The benchmark processing mainly consists of three step, the benchmark's
% internal logic to handle an input, the GPU rendering, and the copy of the
% frame from GPU to the CPU.  The benchmark processing mainly consists of three
% stages, AL, FC and RD.
As GPU rendering (RD) executes in parallel with the application logic (AL) and
frame copy (FC) stages, the GPU rendering times are shown as separate bars in
\FIG~\ref{fig:multi_bench_app_time}.
% The GPU rendering time may be longer or shorter than the application logic
% time, depending on the complexity of the logic and the frames. However,
Surprisingly, many benchmarks spent most of their time on copying frames.
% This long copying time was partially due to the frame copy was delayed until
% the next frame's logic is finished in the graphics pipeline. It is also
This long frame-copy time was due to the long PCIe transporting time and
inefficiency implementation, which were addressed with new optimizations in
Section~\ref{sec:optimization}.
% suggesting this copying time cloud be major optimization target. Later, we
% will show an optimization technique for frame
% copying.% Note that, frame copying did not only require CPU time. In fact, CPU spent a sizable portion of the frame copying time on waiting for the GPU to send the frames over the PCIe bus.
Moreover, because of the long frame-copy, GPU rendering was never the performance
bottleneck in our experiments. 

%\TAB~\ref{tab:latency_breakdown_table} gives the actual time spent on each step of the input processing.

\subsubsection{Architecture-level Resource Usages}
\begin{figure}
  \centering
  \includegraphics[width=8cm]{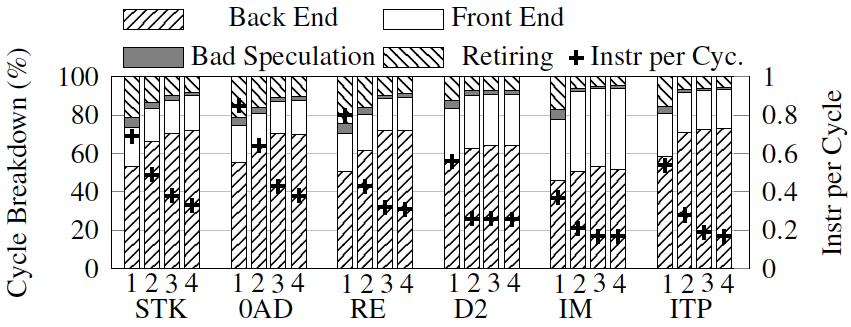}
  \vspace{-4mm}
  \caption{CPU cycles breakdown when executing one to four instances of a
    benchmark on the same server.}
  \label{fig:multi_bench_cpu_cycles_breakdown}
\end{figure}

\begin{figure}
  \centering
  \includegraphics[width=8cm]{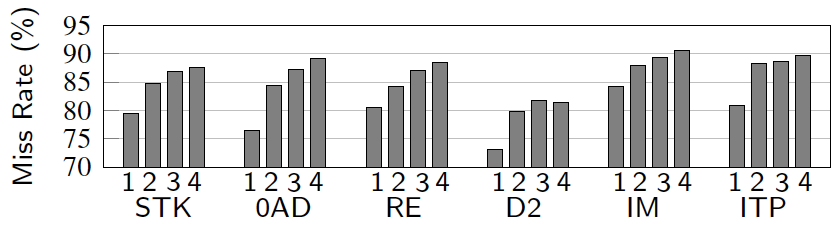}
  \vspace{-4mm}
  \caption{L3 cache miss rates when executing one to four instances of a benchmark on the same server.}
  \label{fig:multi_bench_l3_miss_rates}
  \vspace{-3mm}
\end{figure}

\begin{figure}
  \centering
  \includegraphics[width=8cm]{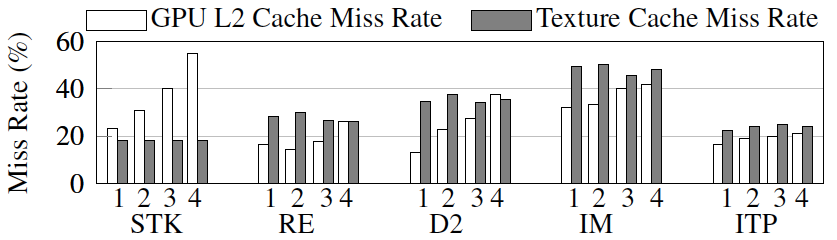}
  \vspace{-4mm}
  \caption{GPU L2 and texture cache miss rates when executing one to four instances of a benchmark.}
  \label{fig:multi_bench_gpu_cache_misses}
\end{figure}

\FIG~\ref{fig:multi_bench_cpu_cycles_breakdown} shows the CPU cycles for each
benchmark using the Top-Down analysis~\cite{intel-opt-manual}. The CPU cycles
are broken down into front-end stalls, back-end stalls, bad speculation stalls,
and the cycles for instruction retirements. As
\FIG~\ref{fig:multi_bench_cpu_cycles_breakdown} shows, all benchmarks had long
back-end stalls and low instructions-per-cycle, indicating these benchmarks were
likely memory-bound. As their L3 cache miss rates (L3-misses/L3-accesses) were
also very high (> 70\%) ( \FIG~\ref{fig:multi_bench_l3_miss_rates}), it can be
deduced that these benchmarks are also off-chip memory bound. This behavior is
consistent with typical graphics rendering implementation, where uncached
memory is used for CPU-to-GPU communications~\cite{intel-writecombine}.

As shown in \FIG~\ref{fig:multi_bench_gpu_cache_misses}, all benchmarks, except
\textit{InMind}, had moderate GPU cache miss rates. These moderate cache miss
rates suggested most benchmarks can use GPU caches relatively effectively.
% and may be sensitive to GPU cache contention.  The related high cache misses
% with \textit{InMind} indicate it may need to optimize its rendering
% implementation.
Note that, \textit{0AD} used OpenGL v1.3, which is not supported by NVidia PMU
reading tools. Therefore, we could not obtain GPU cache miss rates for
\textit{0AD}.

\subsubsection{Single Benchmark Analysis Summary}
1) Executing 3D applications in the cloud can provide reasonable QoS with
current hardware and network.  2) Cloud/server performance can be a major
limitation on FPS and RTT. Therefore, optimizing the cloud system design is
crucial for cloud graphics rendering.  3) 3D applications have a wide range of
resource demands and behaviors, suggesting cloud system optimizations may need
to consider individual application's characteristics.  4) 3D applications
intensively utilized the CPU, GPU, memory and PCIe buses. Consequently, cloud
system optimizations need to consider the impacts of all these resources. For
instance, we designed an optimization to handle the long frame-copy time over
the PCIe bus in Section~\ref{sec:optimization}.

%%% Local Variables:
%%% mode: latex
%%% TeX-master: "paper"
%%% End:

\subsection{Perf. Analysis with Multiple Benchmarks}\label{sec:multi_benchmarks}
To investigate the feasibility and analyze the performance of multiple 3D
applications sharing hardware in the cloud, we also conducted experiments with
multiple 3D benchmarks. More specifically, we executed one to four instances of
the same benchmark on our server. Each benchmark instance interacted with its
own client machine.
% , which had the same configuration as mentioned in
% Section~\ref{sec:evaluation}.
To ensure enough network bandwidth, each benchmark instance used its own 1Gpbs
network card on the server.

\begin{figure}
  \centering
  \includegraphics[width=8cm]{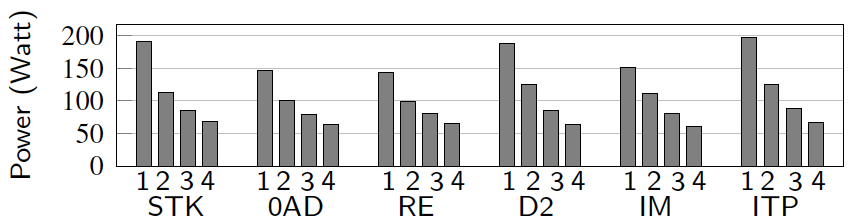}
  \vspace{-4mm}
  \caption{Per-instance power usage when executing one to four benchmark instances on one
    machine.}
  \label{fig:multi_bench_power}
  \vspace{-3mm}
\end{figure}

\subsubsection{Server Power Consumption}
We obtained the server power consumption using a Klein Tools CL110
meter. Overall, adding a new instance only increased the total server power
consumption by less than 20\%.  As shown in \FIG~\ref{fig:multi_bench_power},
This small increase in total power usage translated into per-instance power
usage reductions of 33\%, 50\%, and 61\%, when running two to four instances
(comparing to one instance).
%the power consumption increased by 18\% (\textit{SuperTuxKart}) to 47\%
%(\textit{InMind}), when the instance count was increased from one to
%two. Comparing to one instance, the server power consumption increased by 33\%,
%49\% and 55\% from two to four instances, on average. These results translated
%to
These power usage reductions demonstrate a main benefit of executing 3D
applications in the cloud -- the reduced energy cost and operational cost.

\subsubsection{Application Performance}
\FIG~\ref{fig:multi_bench_fps} also shows the FPS for each benchmark,
when two to four instances of the same benchmark were executed on the same
server.  Here, we focus on the FPS for 2 to 4 instances (i.e., the bars with
x-axis labels ``2'', ``3'' and ``4'').  As \FIG~\ref{fig:multi_bench_fps} shows,
for all benchmarks, executing with two instances could still provide an
acceptable (i.e., $\ge$ 25) FPS. For three benchmarks, \textit{RE}, \textit{IM},
and \textit{ITP}, executing three instances could still achieve an FPS higher
than 25. These FPS results show that consolidating multiple 3D applications on
one server can still provide acceptable QoS,
%can
%indeed reduce the required number of servers while achieving acceptable QoS, and
and thus reduce infrastructure cost.

\FIG~\ref{fig:multi_bench_rtt}, \ref{fig:multi_bench_server_time} and
\ref{fig:multi_bench_app_time} also give the breakdown of the RTT, server
processing time and benchmark processing time, when two to four instances of the
same benchmark were executed on the same server. Again, we focus on the results
for 2 to 4 instances (i.e., the bars with x-axis labels ``2'', ``3'' and
``4''). As \FIG~\ref{fig:multi_bench_rtt} shows, there was no significant
increase in network time due to the use of multiple graphics.

However, there were significant increases in server execution times. As shown in
\FIG~\ref{fig:multi_bench_server_time} and \FIG~\ref{fig:multi_bench_app_time},
nearly every execution stage on the server experienced increased execution time
with more instances. We have observed high increase (up to 96\%) in execution
time for the stages with inter-process communications (IPC), including the
stages of PS and AS. There were also significantly increase in the stages that
do computations on the CPU and GPU, including the stages of AL, FC, RD, and
CP. In particular, the average application logic (stage AL) time increased by
235\% when executing with four instances, and the average GPU rendering (stage
RD) time increased by 133\% when executing with four instances. These increased
execution time on CPU and GPU were main caused by two issues -- the
oversubscribed CPU/GPU when executing three or four instances, and the hardware
resource contention that happened in the CPU, GPU and PCIe buses. This contention is
discussed in detail in the following section.

\subsubsection{Architecture-level Resource Usages}
\FIG~\ref{fig:multi_bench_cpu_cycles_breakdown} and
\FIG~\ref{fig:multi_bench_l3_miss_rates} gives the CPU cycle breakdown and the
L3 cache miss rates of one benchmark instance, when it executed with other
benchmark instances. As the figures show, both the back-end stalls and the L3
miss rates increased considerably with more benchmark instances, indicating that
there was heavy contention in the memory system.

Memory contention was also observed in the GPU. As shown in
\FIG~\ref{fig:multi_bench_gpu_cache_misses}, all benchmarks experienced
increased GPU L2 miss rates, which contributed to the increase in their GPU
rendering time. This L2 miss increases may be explained with the GPU internal
graphics pipelines~\cite{2007-Luebke-IEEEComputer}. Because of this pipeline,
there may be frames from different benchmark instances rendered simultaneously,
thus causing the L2 cache contention. The texture cache miss rates, however, did
not change significantly, as it is a private cache. Note that, for both CPU and
GPU, the contention may exist beyond cache and extend to DRAM and PCIe
buses. Resource contention also existed between the benchmarks and VNC
proxies. However, a full contention analysis is beyond the scope of this paper
and will be conducted in the future.

\subsubsection{Multi-benchmark Analysis Summary}
1) Executing multiple 3D applications on the same server in the cloud can
provide acceptable QoS while significantly reduced energy
consumption. Therefore, cloud graphics rendering may considerably reduce the
infrastructure and operational costs for the large-scale deployment of 3D
applications.  2) Resource contention and slowed IPC can severely degrade the
performance of 3D applications in the cloud, and thus should be properly
managed. Moreover, resource contention simultaneously exists in the CPU and GPU
(and potentially in the PCIe buses). Contention also exists between the
applications and the server proxies. Therefore, resource contention and IPC
management for cloud graphics rendering should be designed with heterogeneity in
mind.

%%% Local Variables:
%%% mode: latex
%%% TeX-master: "paper"
%%% End:

\subsection{Perf. Analysis with Mixed Benchmarks}\label{sec:mixed_benchmarks}
\begin{figure}
  \centering
  \includegraphics[width=8cm]{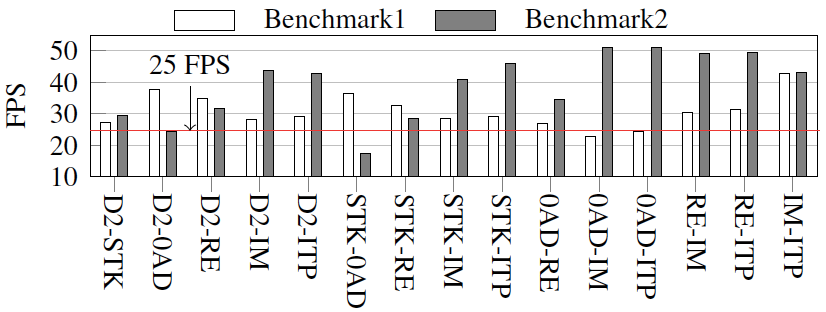}
  \vspace{-4mm}
  \caption{Client FPS for 15 pairs of benchmarks.}
  \label{fig:mixed_bench_fps}
\end{figure}

\begin{figure}
  \centering
  \includegraphics[width=8cm]{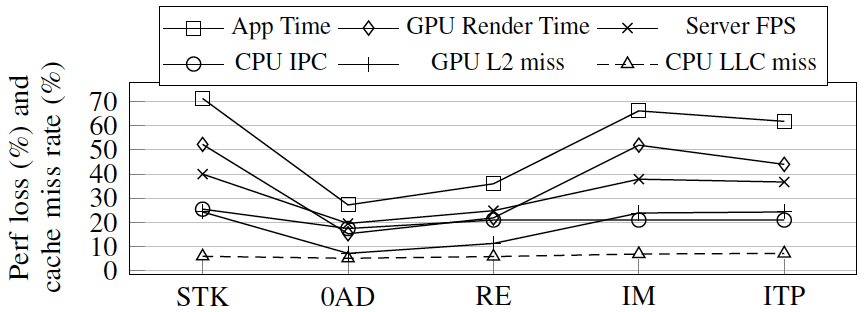}
  \vspace{-4mm}
  \caption{Performance loss and cache miss increases of \textit{Dota2} when
    executing with other benchmarks. Higher values indicate higher loss and
    contention.}
  \vspace{-7mm}
  \label{fig:mixed_bench_dota}
\end{figure}

To study the impact of colocating different 3D applications, we also conducted
experiments where two different benchmarks were executed simultaneously. As
there were 6 benchmarks, a total of 15 pairs of them were evaluated.

\subsubsection{Application Performance, Power Consumption and
  Architectural-level Resource Usages}
\FIG~\ref{fig:mixed_bench_fps} gives the client FPS of the 15 pairs of
benchmarks. Server FPS was just slightly higher than client FPS and was omitted
due to space limitation. As \FIG~\ref{fig:mixed_bench_fps} shows, 11 pairs of
benchmarks had client FPS higher than 25, suggesting different 3D applications
can also share hardware while ensuring acceptable QoS. We also observed that
adding an additional benchmark only increase the total server energy consumption
by no more than 25\%. Therefore, comparing to running two applications on two
servers, executing two different 3D applications on the same server can reduce
energy consumption by at least 37\%.

Similar to the observations in Section~\ref{sec:multi_benchmarks}, oversubscribed
CPU/GPU, prolonged IPC, and resource contention significantly increased the
server execution time. %(including the VNC proxy time,
% application execution time and IPC time)
Nonetheless, we also observe that the contentiousness of these benchmarks varies
considerably. \FIG~\ref{fig:mixed_bench_dota} gives the performance and CPU/GPU
cache misses of \textit{Dota2} when it was executed with different
benchmarks. Other benchmark pairs showed similar results and were omitted due to
space limitation. As \FIG~\ref{fig:mixed_bench_dota} shows, there was
significant variation in \textit{Dota2}'s performance depending on its
co-runners, with \textit{SuperTuxKart} causing the highest contention and
\textit{0AD} causing the least contention. This high variation in the
contentiousness may be utilized in optimizations (e.g., selecting the proper set
of 3D applications to share hardware). It is also interesting to observe that
the contentiousness for CPU cache and GPU cache seemed to have high
correlation. This correlation may be due to the rendering data being shared
between CPU and GPU, and may be exploited when managing 3D applications
contention (e.g., predicting a 3D application's contentiousness).

\subsubsection{Mixed Benchmark Analysis Summary}
1) Executing multiple different 3D applications on the same server in the cloud
can provide acceptable QoS while significantly reducing energy consumption.  2)
The contentiousness of 3D applications varies considerably, which may be
utilized in system optimizations. 3) There may also be a correlation between the
contentiousness for the CPU cache and GPU cache. which may also be exploited in
system optimizations.

%%% Local Variables:
%%% mode: latex
%%% TeX-master: "paper"
%%% End:
\subsection{Container Overhead}\label{sec:dock_overhead}
\begin{figure}
  \centering
  \includegraphics[width=8cm]{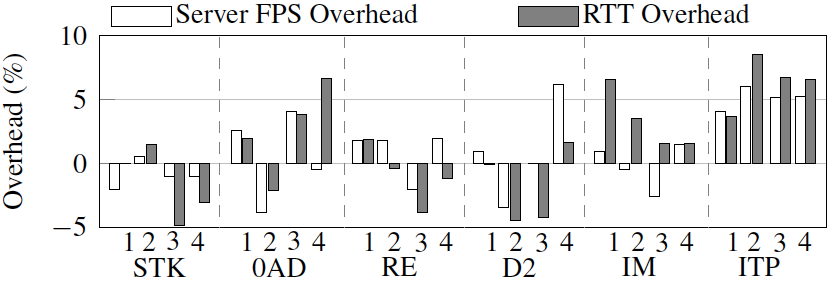}
  \vspace{-4mm}
  \caption{Server FPS/RTT overheads of containers. Negative overheads are
    performance improvements.}
  \label{fig:docker_overhead}
  \vspace{-3mm}
\end{figure}

So far, all experiments were conducted on bare-metal systems. However, as we
target cloud computing, it is also important to analyze the performance impact
of virtualization and containerization. Hence, we repeated the above the
performance analysis experiments using Docker containers to study the overhead
of containerization. More specifically, we executed each instance of the
benchmark and its VNC server inside an NVidia Docker
container~\cite{2018-nvidia-docker}. We chose container instead of VM because
docker container currently supports most GPUs, while VM-based virtualization
require special GPUs. We will evaluate VM-based systems in the future.

\FIG~\ref{fig:docker_overhead} shows the percentages of reduced FPS and
increased RTT (i.e., FPS and RTT overheads), comparing to the runs without
virtualization. On average, Docker containers incurred little overhead. The
average overhead for RTT was only 1.3\%, and the average overhead for server FPS
was only 1.5\%. This low average overhead further shows the feasibility of
executing 3D applications in the cloud.

Nonetheless, the overhead can still be as high as 8.5\% for RTT (or 6\% for
FPS). We observed that these overheads were usually due to increased execution
time for the stages with IPC (stages PS and AS). These high overheads show the
need to optimize containerized cloud 3D applications to ensure that worst case
performance still meets QoS goals. Besides the overheads for RTT and FPS, the
GPU rendering time was also increased by 2.9\% on average and 8\% on maximum,
illustrating the overhead of GPU virtualization with containers.

It is also worth noting that container also improved performance in certain
cases, as shown with the negative overheads in
\FIG~\ref{fig:docker_overhead}. A preliminary analysis showed that these
performance improvements were mainly due to containerization reduced resource
contention among the benchmarks and VNC servers. Although further analysis is
still required to identify the exact cause of the reduced contention, these
performance improvements illustrate the potential benefits of container-based
run-time optimizations.

\textbf{Container Overhead Summary} 1) On average, Docker containers incur
limited overhead, further showing the feasibility of executing 3D applications
in the cloud. 2) Nonetheless, high performance overhead may still be observed in
certain cases, suggesting that container overhead reduction is still
required.
% to ensure worst-case performance still meets QoS goals.
3) Container overheads are mainly associated with IPC and GPU virtualization. 4)
Containerization may also improve performance, suggesting the potential of
additional run-time optimizations.

%%% Local Variables:
%%% mode: latex
%%% TeX-master: "paper"
%%% End:

%%% Local Variables:
%%% mode: latex
%%% TeX-master: "paper"
%%% End:
\section{Optimized Frame Copy}\label{sec:optimization}
\begin{figure}
  \centering
  \includegraphics[width=8cm]{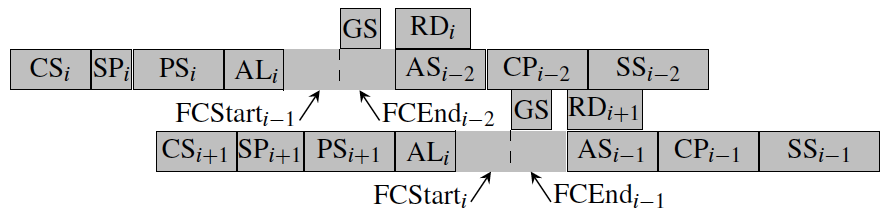}
  \vspace{-4mm}
  \caption{Optimizing frame copy with two-step copy.}
  \label{fig:fast_frame_copy}
  \vspace{-3mm}
\end{figure}

As discussed in Section~\ref{sec:single_bench_app_perf}, the frame-copy (FC)
stage was a major performance bottleneck in TurboVNC. This section presents the
optimizations we invented and implemented to reduce the frame-copying time.

\begin{table*}[h]
  \fontsize{7}{8}\selectfont
  
\def\checkmark{\tikz\fill[scale=0.2](0,.35) -- (.25,0) -- (1,.7) -- (.25,.15) -- cycle;}

\begin{tabular}{|l|c|c|c|c|c|c|c|c|}
  \hline
  Features                      & VNCPlay~\cite{2005-Zeldovich-ATC}  & Chen et al.~\cite{2014-Chen-TOM}   & Slow-Motion~\cite{2003-Nieh-TOCS}  & Login-VSI~\cite{2010-Spruijt-LoginVSI} & DeskBench~\cite{2009-Rhee-INM}     & VDBench~\cite{2010-Berryman-CloudCom} & Dusi et al.~\cite{2012-Dusi-COM}   & \CGraph                            \\ \hline
  Random UI Objects Tolerant    &                                    &                                    &                                    &                                        &                                    &                                       &                                    & \cellcolor[HTML]{32CB00}\checkmark \\ \hline
  Varying Net Latency Tolerant  & \cellcolor[HTML]{32CB00}\checkmark &  &                                    &                                        & \cellcolor[HTML]{32CB00}\checkmark & \cellcolor[HTML]{32CB00}\checkmark    &                                    & \cellcolor[HTML]{32CB00}\checkmark \\ \hline
  User-input Tracking           &                                    &                                    & \cellcolor[HTML]{32CB00}\checkmark &                                        &                                    & \cellcolor[HTML]{32CB00}\checkmark    &                                    & \cellcolor[HTML]{32CB00}\checkmark \\ \hline
  CPU Perf. Measurement       & \cellcolor[HTML]{32CB00}\checkmark & \cellcolor[HTML]{32CB00}\checkmark & \cellcolor[HTML]{32CB00}\checkmark & \cellcolor[HTML]{32CB00}\checkmark     & \cellcolor[HTML]{32CB00}\checkmark & \cellcolor[HTML]{32CB00}\checkmark    & \cellcolor[HTML]{32CB00}\checkmark & \cellcolor[HTML]{32CB00}\checkmark \\ \hline
  Network Perf. Measurement       & \cellcolor[HTML]{32CB00}\checkmark & \cellcolor[HTML]{32CB00}\checkmark & \cellcolor[HTML]{32CB00}\checkmark & \cellcolor[HTML]{32CB00}\checkmark     & \cellcolor[HTML]{32CB00}\checkmark & \cellcolor[HTML]{32CB00}\checkmark    & \cellcolor[HTML]{32CB00}\checkmark & \cellcolor[HTML]{32CB00}\checkmark \\ \hline
  GPU Perf. Measurement       &                                    &                                    &                                    &                                        &                                    &                                       &                                    & \cellcolor[HTML]{32CB00}\checkmark \\ \hline
  PCIe frame-copy Perf. Measure. &                                    &                                    &                                    &                                        &                                    &                                       &                                    & \cellcolor[HTML]{32CB00}\checkmark \\ \hline
  Unaltered 3D App Behaviors       & \cellcolor[HTML]{32CB00}\checkmark &                                    &                                    & \cellcolor[HTML]{32CB00}\checkmark     & \cellcolor[HTML]{32CB00}\checkmark &                                       & \cellcolor[HTML]{32CB00}\checkmark & \cellcolor[HTML]{32CB00}\checkmark \\ \hline
\end{tabular}

%%% Local Variables: 
%%% mode: latex
%%% TeX-master: "paper.tex"
%%% End:
  \caption{Comparison between \CGraph and prior work on VDI and cloud gaming
    performance analysis.}
  \label{tab:features}
\end{table*}

Further analysis of the TurboVNC's graphics interposer revealed two
inefficiencies. First, the interposer called the function
\textit{XGetWindowAttributes} before copying a frame.
\textit{XGetWindowAttributes} was extremely slow and consumed 6\textasciitilde
9ms. This function was only used to get the benchmark's resolution to determine
the size of the frame to copy. As the resolution of a game or VR application is
rarely changed during execution, there is no need to call
\textit{XGetWindowAttributes} for every frame copy. Therefore, in the first
optimization, we applied memoization to this function. That is, we intercepted
the invocation to \textit{XGetWindowAttributes} and returned the cached
resolution instead of actually calling it.  \textit{XGetWindowAttributes} is
only actually invoked when the benchmark's resolution changes, which is
determined by monitoring X events at Hook$_4$.

The second inefficiency is that the benchmarks were halted during the copy,
waiting from the GPU to send the frame, as shown with the blank in the FC stage
in \FIG~\ref{fig:graphics_pipeline}. Inspired by deep CPU pipelining, we broke
the frame copy into two smaller steps -- the start-copy and finish-copy. As
shown in \FIG~\ref{fig:fast_frame_copy}, after issuing the frame copying command
to GPU for frame$_{i-1}$ (FCStart$_{i-1}$), the graphics interposer does not
wait for frame$_{i-1}$'s copy to finish. Instead, it goes on to finish the
copying of frame$_{i-2}$ (FCEnd$_{i-2}$) and works on sending frame$_{i-2}$ to
the VNC server. The actual finish of copying frame$_{i-1}$ happens after the
application logic for frame$_{i+1}$ is computed (FCEnd$_{i-1}$). By making the
frame copy into two asynchronous steps, the halt in the benchmark is
removed. 

\begin{figure}
  \centering
  \includegraphics[width=8cm]{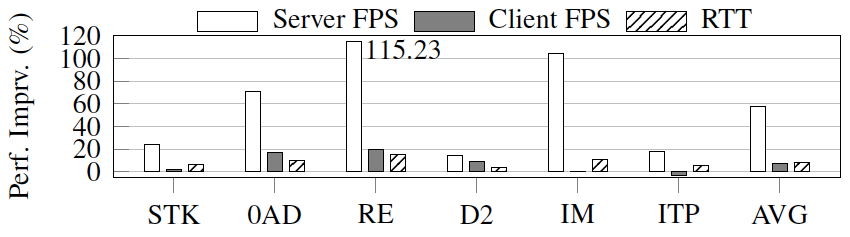}
  \vspace{-4mm}
  \caption{Improved FPS/RTT with our optimizations.}
  \label{fig:optimized_fps}
  \vspace{-3mm}
\end{figure}

\FIG~\ref{fig:optimized_fps} gives the performance improvements from our two
optimizations when one benchmark instance was executed. Our optimization
improved server FPS by 57.7\% on average and 115.2\% at maximum. The client FPS
was improved by 7.4\% on average and 19.5\% at maximum. The RTT was reduced by
8.5\% on average and 15.1\% at maximum. Note that, in
\FIG~\ref{fig:optimized_fps}, the client FPS of \textit{ITP} had 3\% reduction
due to the increased benchmark performance causing more contention with the VNC
proxy. We were able to remove this extra contention and improve \textit{ITP}'s
client FPS with an additional optimization. However, due to space limitation,
this additional optimization cannot be covered here.

%%% Local Variables:
%%% mode: latex
%%% TeX-master: "paper.tex"
%%% End:
\section{Related Work}\label{sec:related}

\textbf{VDI and Cloud Gaming System Benchmarking} There have been several
studies providing benchmarking tools or methodologies to analyze the performance
of VDI systems and cloud gaming systems. \TAB~\ref{tab:features} compares the
functionalities and features of the major prior work with \CGraph. All of the
studies summarized in \TAB~\ref{tab:features}, expect Chen et
al.~\cite{2014-Chen-TOM}, were designed to measure VDI systems with 2D
applications. Moreover, none of these studies considered the random and
irregular UI objects in cloud games and VR applications. And none of these
studies provided methods to measure the performance of GPU rendering and
frame-copy over the PCIe connections. These studies also did not provided means
to track input processing without changing the 3D application's resource usage
and behavior, as shown in Section~\ref{sec:evaluation}.
% Moreover, although the study conducted by Chen et al.~\cite{2014-Chen-TOM}
% targeted 3D cloud games, it also did not provide means to measure GPU and
% frame-copy latency. The application time measured in this work were also
% conducted without remote 3D rendering.  This work also did not intend to
% simulate human actions, and hence, could not handle the random UI objects.
% Furthermore, except for Slow-Motion~\cite{2003-Nieh-TOCS} and
% VDBench~\cite{2010-Berryman-CloudCom}, none of these studies were able to track
% and associate an input with its response frame. Both Slow-Motion and VDBench
% injected artificial delays to allow only one frame being rendered at a time,
% making it easier to associate an input with a response frame. However, this
% delay changed the application behavior and lead to inaccurate performance
% measurements, as shown in Section~\ref{sec:evaluation}.
In summary, without the abilities to handle irregular and random objects,
associate an input and its response, and measure the GPU and frame-copy
performance, existing benchmarking tools/methodologies cannot provide reliable
and effective performance measurements for cloud 3D rendering systems.
  
\noindent\textbf{GPU Benchmarks}. %There are also GPU benchmarks for none GUI applications. Rodinia is a GPU benchmark suite with GPGPU applications~\cite{2009-Che-IISWC}.
There are also many GPU benchmarks, such as
GraalBench~\cite{2004-Antochi-LCTES}, SPECviewperf~\cite{SPECViewperf},
GFXBench~\cite{GFXBench} Rodinia~\cite{2009-Che-IISWC} and
MGMark~\cite{2018-Sun-CoRR}.
% GraalBench is a mobile 3D benchmark suite with OpenGL API replays~\cite{2004-Antochi-LCTES}.  and GFXBench are two popular 3D rendering benchmark suites~\cite{SPECViewperf}. MGMark is a multiple-GPU benchmark suite with GPGPU applications~\cite{2018-Sun-CoRR}.
Mitra and Chiueh also analyzed three 3D benchmark
suites~\cite{1999-Mitra-Micro}.  These benchmarks and analysis focused on
evaluating GPU performance without user actions. However, as 3D applications'
behaviors are heavily affected by user actions, user inputs must be considered
in cloud 3D benchmarks to ensure realistic
results. %Considering user inputs is also required to evaluate the RTT for input processing, which is critical to user experience on cloud graphics rendering system.
Moreover, interactive 3D applications have intensive usage for both CPU and
GPU. Only focusing on GPU cannot provide the insights needed to manage
interactive 3D applications and their use of heterogeneous hardware.

\noindent\textbf{Graphics Rendering Software}. Many software supports remote
desktops, such as VNC, NX and
THINC~\cite{1998-Richarson-IC,1986-Scheifler-TOG,2003-Pinzari-NX,2005-Baratto-SOSP}. These
remote desktops usually do not support 3D applications by default. Additionally,
cloud graphics rendering also requires additional support on GPU virtualization
and management for co-running 3D applications. Consequently, additional research
is required to efficiently support 3D applications in cloud.
% To execute 3D applications properly, VNC, X-Window and NX all require the graphics interposer that was analyzed in our paper.
%There were also designs to improve VNC performance on Linux with complete server-side graphics rendering~\cite{2007-Commander-virtualgl,2003-Stegmaier-ISPA}.
%Hwang and Wood studied the scheduling issues of VDI
%applications~\cite{2012-Hwang-IWQoS}.
CloudVR and Furion were two programming
frameworks to support cloud VR~\cite{2017-Lai-MobiCom,2018-Kamarainen-MM}.
%Crawford and O'Boyle evaluated the impact of compiler options on Vulkan-based
%3D applications~\cite{2018-Crawford-ISPASS}.
%Zyulkyarov et al. investigated using transactional memory on game
%servers~\cite{2009-Zyulkyarov-PPoPP}.
Abe et al. employed data prefetching to speed up the cloud processing time for
interactive applications~\cite{2013-Abe-SoCC}.  Ha et al. investigated the
impact of consolidating multimedia and machine-learning applications in the
cloud~\cite{2013-Ha-IC2E}.  AppStreamer dynamically predicted and downloaded
useful portions of a game to mobile devices~\cite{2020-Theera-EWSN}.  Our
research is inspired by these studies and aims at facilitating these graphics
system design studies. Google, Microsoft, NVidia and PARSEC are also building
their proprietary cloud gaming systems, whose designs may be different than the
system analyzed in this paper~\cite{GeforceNow,Stadia,XCloud,PARSECGaming}. \CGraph aims at
facilitating the public research on cloud graphics rendering, so that
open-source solutions can be as good as proprietary solutions. We will
constantly update \CGraph to follow the advances in these open-source solutions.

\textbf{Other Related Work} GUI testing
frameworks~\cite{2013-Choi-OOPSLA,2016-Mao-ISSTA,2014-Lin-FSE,2012-Amalfitano-ASE,2015-Amalfitano-SOFT}
may also be used for benchmarking remotely-rendered applications.  However,
these GUI testing frameworks were not designed for 3D applications with
irregularly-shaped and random UI objects. Google DeepMind and OpenAI Five have
also built AI bots to play
games~\cite{2019-Jaderberg-Science-GameBot,OpenAIFive}. These bots were built to
compete with human. Therefore, their execution may require thousands of
processors~\cite{OpenAI_dota}. Additionally, the AI models used by these bots
required complex training processes for new games, and existing model are not
publicly available. Therefore, these AI bots are not suitable for 3D application
benchmarking. %\CGraph, however, offers a simpler but reliable methodology to
% simulate human actions for 3D applications.
Moreover, prior research on non-cloud VR architecture and
systems~\cite{2019-Xie-HPCA,2019-Anglada-HPCA,2017-Lai-MobiCom,2019-Leng-ISCA}
may also benefit from \CGraph's intelligent clients and benchmarks.

%%% Local Variables: 
%%% mode: latex
%%% TeX-master: "paper.tex"
%%% End:
\section{Conclusion}\label{sec:conclusion}
This paper presents \CGraph, a benchmarking framework for cloud 3D applications
and systems. \CGraph includes an intelligent client to mimic human interactions
with 3D applications with 1.6\% error, and a performance analysis framework
that provides detailed performance measurements for cloud graphics rendering
systems. With \CGraph, we designed a benchmark suite with six 3D
benchmarks. Using these benchmarks, we characterized a current cloud graphics
rendering system and cloud 3D applications, which also showed benefits of cloud
graphics rendering. We also designed new optimizations with \CGraph to address
two newly-found
bottlenecks. %which could improve the average frame rate by 28\%.

%%% Local Variables:
%%% mode: latex
%%% TeX-master: "paper.tex"
%%% End:

% \bibliographystyle{plain}
\bibliographystyle{IEEEtranS}
\bibliography{bibfiles/cloud_rendering,bibfiles/cloud,bibfiles/wei,bibfiles/memory}

\end{document}